\pdfoutput=1
\documentclass[3p,american,citeautoscript,number,pdftex,sort&compress,twocolumn,%
preprint]{elsarticle}
\usepackage{amsfonts,amsmath,amssymb}
\usepackage[T1]{fontenc}
\usepackage{graphicx}
\usepackage[utf8]{inputenc}
\usepackage{microtype}
\usepackage[textsize=footnotesize,obeyFinal]{todonotes}
\usepackage{xspace}
\usepackage{hyperref}
\usepackage[todos]{CMImacros}

\journal{Chemical Physics Letters}

\makeatletter
\def\ps@pprintTitle{%
 \let\@oddhead\@empty
 \let\@evenhead\@empty
 \def\@oddfoot{}%
 \let\@evenfoot\@oddfoot}
\makeatother

\begin{document}
\begin{frontmatter}
\title{Laser-induced acoustic desorption of thermally stable and unstable biomolecules}
\author[cfeldesy,uhhphys]{Zhipeng Huang}
\author[cfeldesy,uhhcui]{Daniel A. Horke}
\author[cfeldesy,uhhphys,uhhcui]{Jochen Küpper\corref{mycorrespondingauthor}}
\cortext[mycorrespondingauthor]{Corresponding author}
\ead{jochen.kuepper@cfel.de}
\ead[url]{https://www.controlled-molecule-imaging.org}

\address[cfeldesy]{Center for Free-Electron Laser Science, Deutsches Elektronen-Synchrotron DESY,
   Notkestrasse 85, 22607 Hamburg, Germany}%
\address[uhhphys]{Department of Physics, Universität Hamburg, Luruper Chaussee 149, 22761 Hamburg,
   Germany}%
\address[uhhcui]{The Hamburg Center for Ultrafast Imaging, Universität Hamburg, Luruper Chaussee
   149, 22761 Hamburg, Germany}

\begin{abstract}
   We evaluated the effect of the laser-induced acoustic desorption (LIAD) process on thermally
   stable and unstable biomolecules. We found that the thermally labile glycine molecule fragmented
   following desorption via LIAD, due to the production of hot molecules from the LIAD process. We
   furthermore observed a rise in translational temperature with increasing desorption laser
   intensity, while the forward velocity was invariant with respect to the desorption laser
   intensity for both glycine and adenine molecules. The forward kinetic energy was in the range of
   the surface stress energy, which supports the previously proposed stress-induced desorption model
   for the laser-induced acoustic desorption process.

\end{abstract}

\begin{keyword}
   Laser-induced acoustic desorption\sep gas-phase\sep biomolecules \sep mass spectrometry
\end{keyword}
\end{frontmatter}

\section{Introduction}
\label{sec:introduction}
Laser-induced acoustic desorption~\cite{Lindner:AnalChem57:895} is a promising technique to bring
thermally labile, light sensitive and non-volatile molecules into gas phase. It relies on samples
being deposited as a thin layer on a metal foil, typical foil thickness of around 10~\um, which are
then desorbed by irradiating the backside of the foil, \ie, the side without sample, with a
nanosecond laser. As this method avoids direct contact between the desorption laser and sample, it
is especially suitable for light-sensitive and labile samples and has been demonstrated for bringing
biological systems ranging from amino acids~\cite{Calvert:PCCP14:6289, Belshaw:JPCL3:3751,
   Calegari:Science346:336, Huang:AnalChem90:3920}, through peptides~\cite{Shea:AnalChem79:2688,
   Shea:AnalChem78:6133, Habicht:AnalChem82:608} and even entire viruses, bacteria and
cells~\cite{Peng:ACIE45:1423, Zhang:AnalChem88:5958} into the gas-phase. Such LIAD-based molecule
sources have been applied to mass spectrometry studies~\cite{Peng:ACIE45:1423,
   Golovlev:ijmsip169:69, Dow:EJMS18:77, Nyadong:AnalChem84:7131}, gas-phase chemical
reactions~\cite{Shea:AnalChem79:1825, Demarais:APJ784:25}, and even attosecond dynamics
experiments~\cite{Calvert:PCCP14:6289, Belshaw:JPCL3:3751, Calegari:Science346:336}. They are
further promising large-molecule sources for use in matter-wave
interferometry~\cite{Sezer:Analchem87:5614, Sezer:JMS50:235} and single-particle imaging experiments
at free-electron lasers~\cite{Barty:ARPC64:415, Kuepper:PRL112:083002}.

We have previously demonstrated and characterized our newly designed LIAD source, featuring constant
sample replenishment using a tape-drive to deliver fresh sample, and prepared a high-density plume
of phenylalanine~\cite{Huang:AnalChem90:3920}. We observed that increasing the pulse energy of the
desorption laser leads to increased fragmentation, as well as a significant increase in the
translational temperature of desorbed molecules. In this contribution we investigated how the LIAD
source parameters, such as desorption laser intensity and desorption-ionization timing, affected the
produced molecular plume for thermally stable and thermally unstable biological molecules, using
adenine and glycine as prototypical examples~\cite{Wang:JAAP111:1, Yablokov:RJGC:79:1704}. Our
results confirmed the previous assignment of a desorption model based on surface stress on sample
islands deposited on the foil, as evidenced by the combination of an invariance of the molecular
plume velocity on the desorption laser intensity, but an increase of the translational
temperature~\cite{Huang:AnalChem90:3920, Zinovev:AnalChem79:8232}. For the thermally labile glycine
sample we found significant further fragmentation during the propagation in the vacuum chamber
following desorption. These molecules were found to possess translational temperatures well above
the decomposition threshold. If there is a full thermalization of internal and external degrees of
freedom, which seems reasonable given the involved microsecond timescale desorption process, this
would explain the observed fragmentation.

\section{Experimental method}
\label{sec:Experimental method}
Detailed descriptions of our experimental setup and sample preparation method were given
before~\cite{Huang:AnalChem90:3920}. Briefly, samples were deposited on the front surface of a
10~\um thick, 10~mm wide and 1~m long tantalum foil band. In order to create a stable coverage of
sample on the foil, sample was aerosolized using a gas-dynamic virtual nozzle
(GDVN)~\cite{DePonte:JPD41:195505, Beyerlein:RSI86:125104} and this aerosol deposited on the foil,
where it sticks and rapidly dries out. The surface coverage of adenine and glycine on the foil was
around 600~nmol/cm$^2$ and 400~nmol/cm$^2$, respectively, as determined by weighing foil bands
before and after sample deposition. To desorb molecules, the back surface of the foil was irradiated
by a 355~nm laser pulse with 8~ns duration at a repetition rate of 20~Hz, focused to a spot size of
around 300~\um (FWHM) on the foil. Sample was constantly replenished during operation by forwarding
the foil band with a velocity of $\sim50$~\um/s. Desorbed molecules were ionized using strong-field
ionization (SFI) induced by a focused femtosecond (40~fs) Ti:Sapphire laser focused to a spot size
of around $100~\um$, corresponding to typical field strengths of $4\times10^{13}$~W/cm$^{2}$.
Produced ions were detected by a linear time-of-flight mass spectrometer (TOF-MS).

\section{Results and Discussion}
\label{sec:results}
\begin{figure}
   \centering%
   \includegraphics[width=\linewidth]{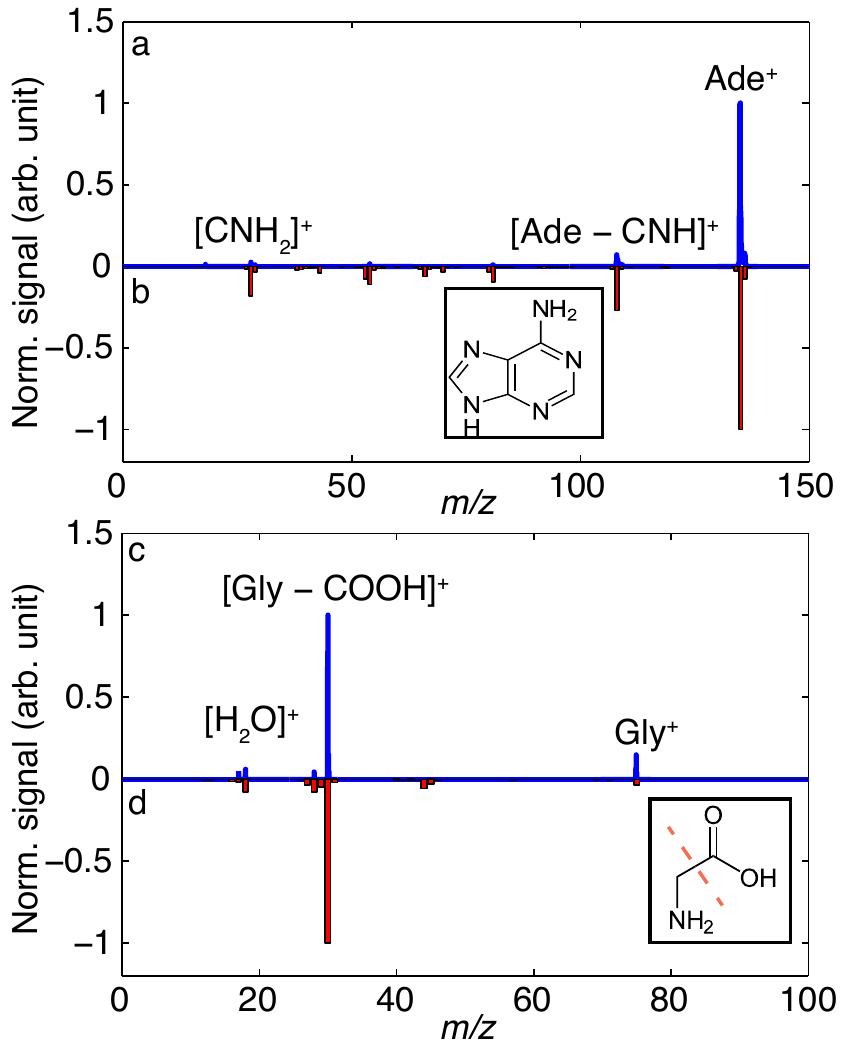}
   \caption{Mass spectrum of adenine and glycine recorded using strong-field ionization with a field
   strengths of $4\times10^{13}$~W/cm$^{2}$(blue) and electron impact ionization
   (red)~\cite{NIST:webbook:2017}. Both are normalized to the dominant mass-to-charge ratio peak.}
   \label{fig:tof}
\end{figure}
Typical time-of-flight mass spectra of adenine and glycine desorbed via LIAD are shown in
\autoref{fig:tof} and are compared to literature spectra obtained using electron-impact ionization
of thermally evaporated samples~\cite{NIST:webbook:2017}. Both spectra were normalized to their
respective dominant ion peak, \ie, the parent ion peak at 135~u for adenine and the dominant
fragment peak at 30~u for glycine. The adenine spectra clearly demonstrate the production of intact
adenine in the gas-phase using LIAD, with very little fragmentation. For glycine, however, strong
fragmentation was observed for both LIAD and the reference spectrum. Nonetheless, using LIAD
combined with SFI a significant contribution from intact glycine was present. The mass spectra
showed no evidence for the formation of molecular clusters or ablation of metal atoms or clusters
from the foil band~\cite{Huang:AnalChem90:3920}.

\begin{figure*}[t]
   \centering%
   \includegraphics[width=0.9\linewidth]{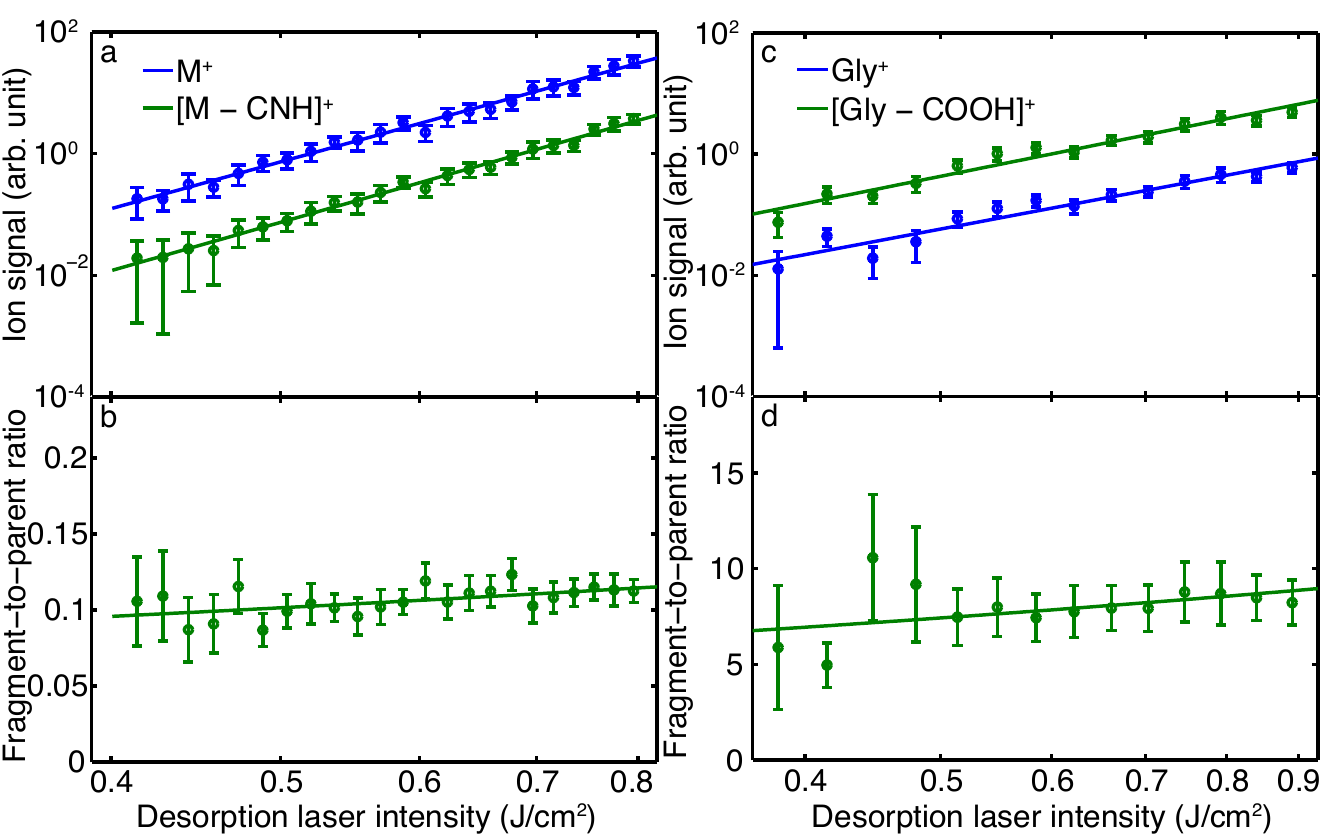}
   \caption{Parent and dominant fragment ion yields (a,c) and fragment-to-parent ratios (b,d) as a
      function of desorption laser intensity for adenine (a,b) and glycine (c,d). Data was recorded
      using strong-field ionization 4.5~mm behind the foil band.}
   \label{fig:nspowerscan}
\end{figure*}
In order to investigate how the desorption laser intensity affects the fragmentation behavior for
thermally stable and labile molecules, respectively, we collected mass spectra for adenine and
glycine under different desorption laser intensities. Molecules were ionized using SFI 4.5~mm behind
the foil band. \autoref[a,c]{fig:nspowerscan} show the respective parent and dominant fragment ion
yields as a function of desorption laser intensity. These data were fit with a power-law dependence
of the form $A\times{}x^n$, and all ion channels showed a corresponding increase with increasing
laser intensity. \autoref[b,d]{fig:nspowerscan} shows the fragment-to-parent ratios for adenine and
glycine as a function of desorption laser intensity and we observed only a slight linear increase in
fragmentation as the desorption power is increased. Both thermally stable and unstable molecules,
therefore, behave similar with increasing desorption laser intensity, confirming the non-thermal
nature of the desorption process.

\begin{figure}
   \centering%
   \includegraphics[width=\linewidth]{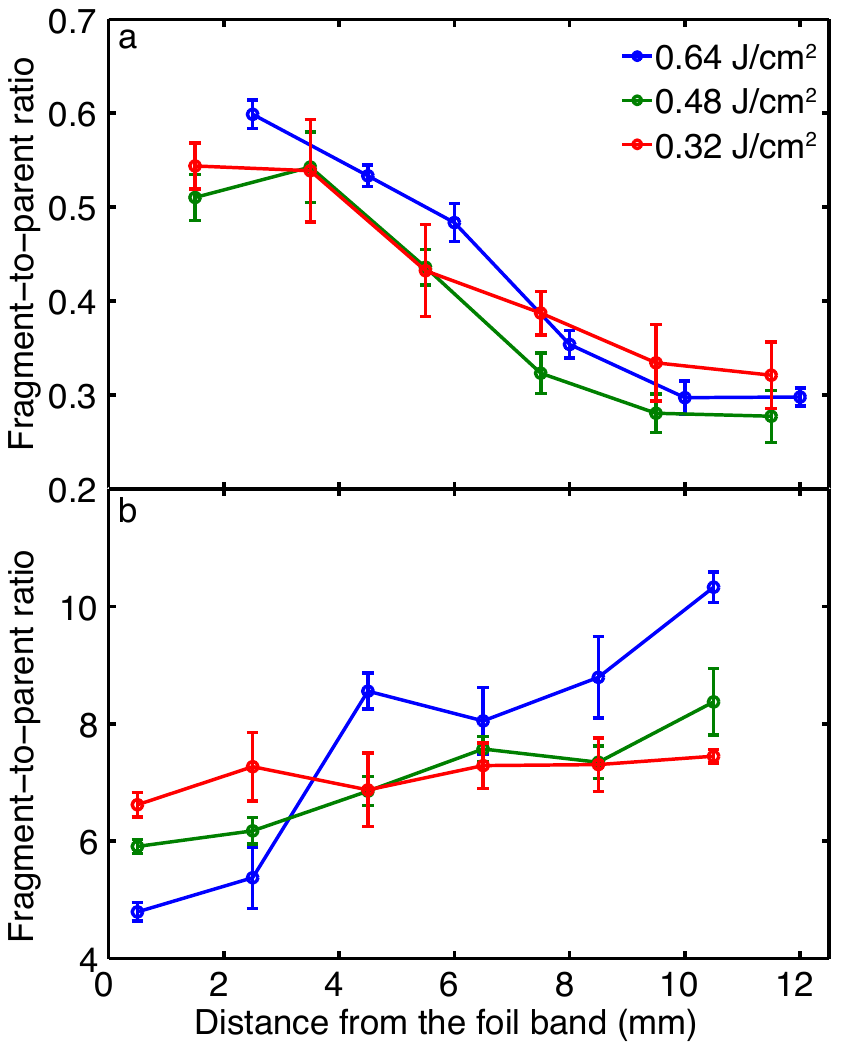}
   \caption{Fragment-to-parent ratio of adenine (a) and glycine (b) as a function of distance from
      the foil band for the highest density part of the plume.}
   \label{fig:fragment}
\end{figure}
To investigate whether further fragmentation occurred after desorption, \ie, during propagation of
the molecular plume through the vacuum chamber, we recorded mass spectra at different distances
behind the foil band. At each distance the delay between ionization laser and desorption laser was
changed, such that we always probed the highest density part of the plume, \ie, we followed the peak
of the plume as it travels. This data is shown in \autoref[a,b]{fig:fragment}, where we plot the
fragment-to-parent ratio for distances of 0.5--12.0~mm between the foil band and the interaction
point. \autoref[a]{fig:fragment} shows the behavior for adenine, using the dominant fragment at
108~u, corresponding to the loss of --CNH from intact adenine. We observed a decrease of the
fragment-to-parent ratio with distance from foil band, indicating that the peak number density of
fragments decreased relatively to the adenine parent. At the same time the absolute densities for
parent and fragment should decrease with increasing distance. We attribute the observed behavior to
different relative velocity distributions for fragments and intact molecules. Fragments appear to be
traveling at higher velocities, likely due to the additional kinetic energy released in the
fragmentation process, leading to the observed decrease in the fragment-to-parent ratio of adenine
with increasing distance from the foil band. This indicates that there was no further fragmentation
of adenine during propagation.

The behavior observed for glycine is markedly different, as shown in \autoref[b]{fig:fragment}. Here, the
relative population of the dominant fragment at 30~u increased during propagation, indicating that
further fragmentation occurred during free flight of the molecules through the chamber. We note that
one might expect fragments and intact glycine to propagate at different velocities, as observed for
adenine, such that the fragmentation during propagation might be even more significant than the data
in \autoref[b]{fig:fragment} suggests. These two competing effects also explain the unclear
variation in a corresponding measurement of the fragmentation during propagation of phenylalanine
following LIAD~\cite{Huang:AnalChem90:3920}. The different behavior for adenine and glycine suggests
a thermal decomposition during free flight propagation as the origin of the enhanced fragmentation.
This will be investigated further below, where we asses the translational temperature of desorbed
molecules.
\begin{figure}
   \centering%
   \includegraphics[width=\linewidth]{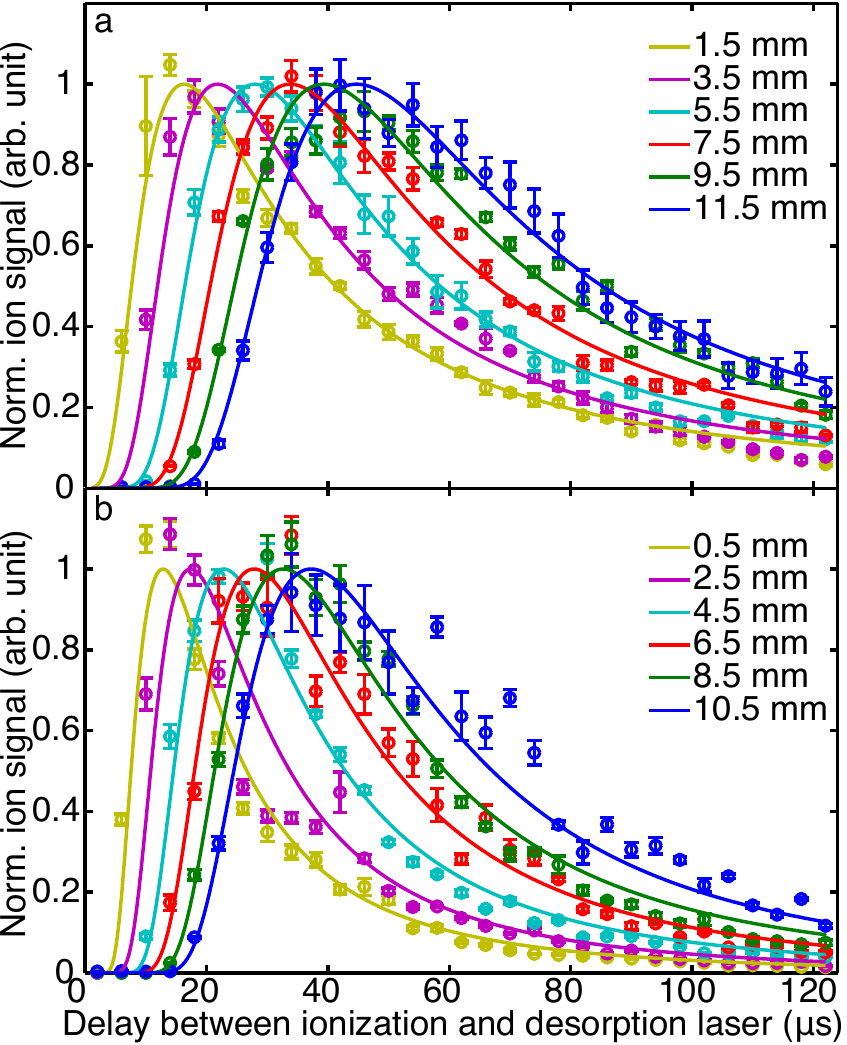}
   \caption{Adenine (a) and glycine (b) plume temporal distribution measured at different distances
      from the foil band for a desorption laser fluence of 0.48~J/cm$^2$.}
   \label{fig:temporal_profiles}
\end{figure}

In order to evaluate the translational forward velocity and translational temperature of the plume,
we recorded mass spectra for different ionization-desorption laser delays and at different distances
from the foil. These data, shown in \autoref{fig:temporal_profiles}, were then modeled by a
Maxwell-Boltzmann distribution convoluted with the initial temporal distribution from the LIAD
process~\cite{Huang:AnalChem90:3920}. The shown data was recorded for a desorption laser fluence of
0.48~J/cm$^2$, profiles for other fluences are shown in the supplementary materials. All
experimental data were fitted globally for all propagation distances with a common temperature $T$
and forward velocity $v_{0,z}$. The resulting fit is shown as solid lines in
\autoref{fig:temporal_profiles}. The extracted forward velocity and translational temperature for
different desorption laser intensities are summarized in \autoref{tab:temperature}. We found that
the forward velocity of both, adenine and glycine, plumes was invariant with respect to desorption
laser intensity, as previously observed for other molecules~\cite{Huang:AnalChem90:3920,
   Zinovev:AnalChem79:8232}. Similarly we confirmed the previous finding that the translational
temperature increases with desorption laser intensity, for both, adenine and glycine. However, while
the former only shows a very modest increase over the laser fluence range investigated, the glycine
translational temperature was found to increase significantly. For all but the lowest desorption
fluence, the extracted translational temperature for the glycine plume was above its thermal
decomposition temperature, which is on the order of $\ordsim485-513$~K~\cite{Yablokov:RJGC:79:1704}.
The translational temperatures for adenine, however, were much below its decomposition temperature
of 582~K~\cite{Wang:JAAP111:1}. While we have no definitive explanation for the differences in
observed translational temperature, we speculate that this is due to different thermal properties,
such as heat transport, from the metal foil to the sample during the desorption process.
Furthermore, we note that the different thicknesses of the sample layers will influence this.
Nonetheless, these results rationalize the observed fragmentation behavior following desorption from
the foil band shown in \autoref{fig:fragment}: the thermally unstable glycine fragmented further,
while for adenine no additional fragmentation was observed.
\begin{table}
   \centering%
   \caption{Derived translational forward velocities $v_{0, z}$ and translational temperatures $T$
      in the moving frame for adenine and glycine at different desorption-laser fluences.}
   \label{tab:temperature}
   \footnotesize%
   \begin{tabular}{*{5}{c}}
     \hline\hline
     Fluence & \multicolumn{2}{c}{Ade} & \multicolumn{2}{c}{Gly}\\
     (J/cm$^2$) & ~$T$~(K)~ & ~$v_{0, z}$~(m/s)~ & ~$T$~(K)~ & ~$v_{0, z}$~(m/s)~ ~\\
     \hline
     0.32 & 494 & 365 & 441 & 337   \\
     0.48 & 516 & 369 & 496 & 338   \\
     0.64 & 521 & 384 & 698 & 355   \\
     0.80 & 523 & 380 & 745 & 340   \\
     \hline\hline
   \end{tabular}
\end{table}

The observation of a constant plume velocity, but increasing translational temperature, as well as
only a modest increase in fragmentation with increasing desorption laser fluence, is a clear
indicator for a desorption process based on surface stress between the substrate and the deposited
sample~\cite{Zinovev:AnalChem79:8232, Huang:AnalChem90:3920}. The forward velocity of the molecular
plume, \ie, its kinetic energy, is hence determined by material properties of the substrate and
sample. From our measurements we extracted a forward kinetic energy, based on the average velocity,
of around 90~meV for adenine and 45~meV for glycine. This is consistent with the previously observed
kinetic energy for phenylalanine under identical conditions, around
50~meV~\cite{Huang:AnalChem90:3920}, and within the range of simple estimates of the stress energy
of 25--100~meV~\cite{Zinovev:AnalChem79:8232, Bondi:JAP37:4643}. The kinetic energy of desorbed
molecules is essentially a measure of the interaction energy between the deposited molecules and the
metal foil. In a very qualitative picture it is easy to rationalize how adenine with its highly
delocalized $\pi$-system has a stronger interaction with a metal surface than glycine, with
phenylalanine somewhere between the two.

\section{Conclusion}
\label{sec:Conclusion}
We demonstrated the use of LIAD for the production of gas-phase samples of adenine and glycine,
prototypical examples of thermally stable and unstable biological molecules, respectively. We showed
that the high translational temperatures of molecules following desorption correlate with further
fragmentation of thermally unstable systems as they travel through the vacuum chamber, indicating a
strong coupling between internal and external degrees of freedom in the desorbed samples.
Measurements of the translational temperature and molecular velocity distributions confirmed this,
and showed that indeed glycine is produced with a temperature in excess of its decomposition
threshold. The invariance of forward velocities on the desorption laser power further confirmed a
desorption model based on surface stress between substrate and sample.

The additional fragmentation of molecules could be avoided if they were rapidly cooled down, for
example using buffer-gas cells~\cite{Hutzler:CR112:4803} or entrainment in cold molecular beams, as
is commonly done in direct laser desorption~\cite{Teschmit:JCP147:144204}. Such experiments are
underway in our laboratory and such a cold beam of intact large molecules would enable novel
experiments, from controlling large molecules with external fields~\cite{Chang:IRPC34:557,
   Teschmit:ACIE57:13775} to direct diffractive imaging of single molecules in the gas-phase~\cite
{Barty:ARPC64:415, Kuepper:PRL112:083002}.

\section*{Acknowledgments}
This work has been supported by the European Research Council under the European Union's Seventh
Framework Programme (FP7/2007-2013) through the Consolidator Grant COMOTION (ERC-614507-Küpper), by
the excellence cluster ``The Hamburg Center for Ultrafast Imaging -- Structure, Dynamics and Control
of Matter at the Atomic Scale'' of the Deutsche Forschungsgemeinschaft (CUI, DFG-EXC1074), and by
the Helmholtz Gemeinschaft through the ``Impuls- und Vernetzungsfond''. Z.\,H.\ gratefully
acknowledges a scholarship of the Joachim-Herz-Stiftung and financial support from the PIER
Helmholtz Graduate School.

\bigskip

\bibliography{LIAD-2.bbl}

\begin{thebibliography}{10}
\expandafter\ifx\csname url\endcsname\relax
  \def\url#1{\texttt{#1}}\fi
\expandafter\ifx\csname urlprefix\endcsname\relax\def\urlprefix{URL }\fi
\expandafter\ifx\csname href\endcsname\relax
  \def\href#1#2{#2} \def\path#1{#1}\fi

\bibitem{Lindner:AnalChem57:895}
B.~Lindner, U.~Seydel,
  \href{http://pubs.acs.org/doi/abs/10.1021/ac00281a027}{Laser desorption mass
  spectrometry of nonvolatiles under shock wave conditions}, Anal.\ Chem. 57
  (1985) 895--899.
\newblock \href {http://dx.doi.org/10.1021/ac00281a027}
  {\path{doi:10.1021/ac00281a027}}.
\newline\urlprefix\url{http://pubs.acs.org/doi/abs/10.1021/ac00281a027}

\bibitem{Calvert:PCCP14:6289}
C.~R. Calvert, L.~Belshaw, M.~J. Duffy, O.~Kelly, R.~B. King, A.~G. Smyth,
  T.~J. Kelly, J.~T. Costello, D.~J. Timson, W.~A. Bryan, T.~Kierspel, P.~Rice,
  I.~C.~E. Turcu, C.~M. Cacho, E.~Springate, I.~D. Williams, J.~B. Greenwood,
  \href{http://dx.doi.org/10.1039/c2cp23840c}{{LIAD}-fs scheme for studies of
  ultrafast laser interactions with gas phase biomolecules}, Phys.\ Chem.\
  Chem.\ Phys. 14~(18) (2012) 6289--6297.
\newblock \href {http://dx.doi.org/10.1039/c2cp23840c}
  {\path{doi:10.1039/c2cp23840c}}.
\newline\urlprefix\url{http://dx.doi.org/10.1039/c2cp23840c}

\bibitem{Belshaw:JPCL3:3751}
L.~Belshaw, F.~Calegari, M.~J. Duffy, A.~Trabattoni, L.~Poletto, M.~Nisoli,
  J.~B. Greenwood,
  \href{http://pubs.acs.org/doi/abs/10.1021/jz3016028}{Observation of ultrafast
  charge migration in an amino acid}, J.\ Phys.\ Chem.\ Lett. 3~(24) (2012)
  3751--3754.
\newblock \href {http://dx.doi.org/10.1021/jz3016028}
  {\path{doi:10.1021/jz3016028}}.
\newline\urlprefix\url{http://pubs.acs.org/doi/abs/10.1021/jz3016028}

\bibitem{Calegari:Science346:336}
F.~Calegari, D.~Ayuso, A.~Trabattoni, L.~Belshaw, S.~De~Camillis, S.~Anumula,
  F.~Frassetto, L.~Poletto, A.~Palacios, P.~Decleva, J.~B. Greenwood,
  F.~Mart{\'\i}n, M.~Nisoli,
  \href{http://science.sciencemag.org/content/346/6207/336.full}{Ultrafast
  electron dynamics in phenylalanine initiated by attosecond pulses}, Science
  346~(6207) (2014) 336--339.
\newblock \href {http://dx.doi.org/10.1126/science.1254061}
  {\path{doi:10.1126/science.1254061}}.
\newline\urlprefix\url{http://science.sciencemag.org/content/346/6207/336.full}

\bibitem{Huang:AnalChem90:3920}
Z.~Huang, T.~Ossenbr{\"u}ggen, I.~Rubinsky, M.~Schust, D.~A. Horke,
  J.~K{\"u}pper,
  \href{http://pubs.acs.org/doi/10.1021/acs.analchem.7b04797}{Development and
  characterization of a laser-induced acoustic desorption source}, Anal.\ Chem.
  90~(6) (2018) 3920--3927.
\newblock \href {http://arxiv.org/abs/1710.06684} {\path{arXiv:1710.06684}},
  \href {http://dx.doi.org/10.1021/acs.analchem.7b04797}
  {\path{doi:10.1021/acs.analchem.7b04797}}.
\newline\urlprefix\url{http://pubs.acs.org/doi/10.1021/acs.analchem.7b04797}

\bibitem{Shea:AnalChem79:2688}
R.~C. Shea, S.~C. Habicht, W.~E. Vaughn, H.~I. Kentt{\"a}maa,
  \href{http://pubs.acs.org/doi/abs/10.1021/ac061597p}{Design and
  characterization of a high-power laser-induced acoustic desorption probe
  coupled with a fourier transform ion cyclotron resonance mass spectrometer},
  Anal.\ Chem. 79~(7) (2007) 2688--2694.
\newblock \href {http://dx.doi.org/10.1021/ac061597p}
  {\path{doi:10.1021/ac061597p}}.
\newline\urlprefix\url{http://pubs.acs.org/doi/abs/10.1021/ac061597p}

\bibitem{Shea:AnalChem78:6133}
R.~C. Shea, C.~J. Petzold, J.~L. Campbell, S.~Li, D.~J. Aaserud, H.~I.
  Kentt{\"a}maa,
  \href{http://pubs.acs.org/doi/abs/10.1021/ac0602827}{Characterization of
  laser-induced acoustic desorption coupled with a fourier transform ion
  cyclotron resonance mass spectrometer}, Anal.\ Chem. 78~(17) (2006)
  6133--6139.
\newblock \href {http://dx.doi.org/10.1021/ac0602827}
  {\path{doi:10.1021/ac0602827}}.
\newline\urlprefix\url{http://pubs.acs.org/doi/abs/10.1021/ac0602827}

\bibitem{Habicht:AnalChem82:608}
S.~C. Habicht, L.~M. Amundson, P.~Duan, N.~R. Vinueza, H.~I. Kentt{\"a}maa,
  \href{http://pubs.acs.org/doi/abs/10.1021/ac901943k}{Laser-induced acoustic
  desorption coupled with a linear quadrupole ion trap mass spectrometer},
  Anal.\ Chem. 82~(2) (2010) 608--614.
\newblock \href {http://dx.doi.org/10.1021/ac901943k}
  {\path{doi:10.1021/ac901943k}}.
\newline\urlprefix\url{http://pubs.acs.org/doi/abs/10.1021/ac901943k}

\bibitem{Peng:ACIE45:1423}
W.-P. Peng, Y.-C. Yang, M.-W. Kang, Y.-K. Tzeng, Z.~Nie, H.-C. Chang, W.~Chang,
  C.-H. Chen,
  \href{http://onlinelibrary.wiley.com/doi/10.1002/anie.200503271/abstract}{Laser-induced
  acoustic desorption mass spectrometry of single bioparticles}, Angew.\ Chem.\
  Int.\ Ed. 45 (2006) 1423--1426.
\newblock \href {http://dx.doi.org/10.1002/anie.200503271}
  {\path{doi:10.1002/anie.200503271}}.
\newline\urlprefix\url{http://onlinelibrary.wiley.com/doi/10.1002/anie.200503271/abstract}

\bibitem{Zhang:AnalChem88:5958}
N.~Zhang, K.~Zhu, C.~Xiong, Y.~Jiang, H.-C. Chang, Z.~Nie,
  \href{http://pubs.acs.org/doi/abs/10.1021/acs.analchem.6b00918}{Mass
  measurement of single intact nanoparticles in a cylindrical ion trap}, Anal.\
  Chem. 88~(11) (2016) 5958--5962.
\newblock \href {http://dx.doi.org/10.1021/acs.analchem.6b00918}
  {\path{doi:10.1021/acs.analchem.6b00918}}.
\newline\urlprefix\url{http://pubs.acs.org/doi/abs/10.1021/acs.analchem.6b00918}

\bibitem{Golovlev:ijmsip169:69}
V.~V. Golovlev, S.~L. Allman, W.~R. Garrett, N.~I. Taranenko, C.~H. Chen,
  \href{http://linkinghub.elsevier.com/retrieve/pii/S0168117697002097}{Laser-induced
  acoustic desorption}, Int.\ J.\ Mass Spectrom.\ Ion Processes 169--170 (1997)
  69--78.
\newblock \href {http://dx.doi.org/10.1016/S0168-1176(97)00209-7}
  {\path{doi:10.1016/S0168-1176(97)00209-7}}.
\newline\urlprefix\url{http://linkinghub.elsevier.com/retrieve/pii/S0168117697002097}

\bibitem{Dow:EJMS18:77}
A.~Dow, A.~Wittrig, H.~Kentt{\"a}maa,
  \href{http://www.impublications.com/content/abstract?code=E18_0077}{Laser-induced
  acoustic desorption (liad) mass spectrometry}, Eur.\ J.\ Mass Spectrom.
  18~(2) (2012) 77--92.
\newblock \href {http://dx.doi.org/10.1255/ejms.1162}
  {\path{doi:10.1255/ejms.1162}}.
\newline\urlprefix\url{http://www.impublications.com/content/abstract?code=E18_0077}

\bibitem{Nyadong:AnalChem84:7131}
L.~Nyadong, J.~P. Quinn, C.~S. Hsu, C.~L. Hendrickson, R.~P. Rodgers, A.~G.
  Marshall, \href{http://pubs.acs.org/doi/abs/10.1021/ac301307p}{Atmospheric
  pressure laser-induced acoustic desorption chemical ionization mass
  spectrometry for analysis of saturated hydrocarbons}, Anal.\ Chem. 84~(16)
  (2012) 7131--7137.
\newblock \href {http://dx.doi.org/10.1021/ac301307p}
  {\path{doi:10.1021/ac301307p}}.
\newline\urlprefix\url{http://pubs.acs.org/doi/abs/10.1021/ac301307p}

\bibitem{Shea:AnalChem79:1825}
R.~C. Shea, C.~J. Petzold, J.-a. Liu, H.~I. Kentt{\"a}maa,
  \href{http://pubs.acs.org/doi/abs/10.1021/ac061596x}{Experimental
  investigations of the internal energy of molecules evaporated via
  laser-induced acoustic desorption into a fourier transform ion cyclotron
  resonance mass spectrometer}, Anal.\ Chem. 79~(5) (2007) 1825--1832.
\newblock \href {http://dx.doi.org/10.1021/ac061596x}
  {\path{doi:10.1021/ac061596x}}.
\newline\urlprefix\url{http://pubs.acs.org/doi/abs/10.1021/ac061596x}

\bibitem{Demarais:APJ784:25}
N.~J. Demarais, Z.~Yang, T.~P. Snow, V.~M. Bierbaum,
  \href{http://stacks.iop.org/0004-637X/784/i=1/a=25?key=crossref.35f76738aeb2c9005ab459833a5c32ae}{Gas-phase
  reactions of polycyclic aromatic hydrocarbon cations and their
  nitrogen-containing analogs with h atoms}, Astrophys.\ J. 784~(1) (2014)
  25--7.
\newblock \href {http://dx.doi.org/10.1088/0004-637X/784/1/25}
  {\path{doi:10.1088/0004-637X/784/1/25}}.
\newline\urlprefix\url{http://stacks.iop.org/0004-637X/784/i=1/a=25?key=crossref.35f76738aeb2c9005ab459833a5c32ae}

\bibitem{Sezer:Analchem87:5614}
U.~Sezer, L.~W{\"o}rner, J.~Horak, L.~Felix, J.~T{\"u}xen, C.~G{\"o}tz,
  A.~Vaziri, M.~Mayor, M.~Arndt,
  \href{http://pubs.acs.org/doi/abs/10.1021/acs.analchem.5b00601}{Laser-induced
  acoustic desorption of natural and functionalized biochromophores}, Anal.\
  Chem. 87~(11) (2015) 5614--5619.
\newblock \href {http://dx.doi.org/10.1021/acs.analchem.5b00601}
  {\path{doi:10.1021/acs.analchem.5b00601}}.
\newline\urlprefix\url{http://pubs.acs.org/doi/abs/10.1021/acs.analchem.5b00601}

\bibitem{Sezer:JMS50:235}
U.~Sezer, P.~Schmid, L.~Felix, M.~Mayor, M.~Arndt,
  \href{http://doi.wiley.com/10.1002/jms.3526}{Stability of high-mass molecular
  libraries: the role of the oligoporphyrin core}, J.\ Mass.\ Spectrom. 50
  (2015) 235--239.
\newblock \href {http://dx.doi.org/10.1002/jms.3526}
  {\path{doi:10.1002/jms.3526}}.
\newline\urlprefix\url{http://doi.wiley.com/10.1002/jms.3526}

\bibitem{Barty:ARPC64:415}
A.~Barty, J.~K{\"u}pper, H.~N. Chapman,
  \href{http://dx.doi.org/10.1146/annurev-physchem-032511-143708}{Molecular
  imaging using x-ray free-electron lasers}, Annu.\ Rev.\ Phys.\ Chem. 64~(1)
  (2013) 415--435.
\newblock \href {http://dx.doi.org/10.1146/annurev-physchem-032511-143708}
  {\path{doi:10.1146/annurev-physchem-032511-143708}}.
\newline\urlprefix\url{http://dx.doi.org/10.1146/annurev-physchem-032511-143708}

\bibitem{Kuepper:PRL112:083002}
J.~K{\"u}pper, S.~Stern, L.~Holmegaard, F.~Filsinger, A.~Rouz\'{e}e,
  A.~Rudenko, P.~Johnsson, A.~V. Martin, M.~Adolph, A.~Aquila, S.~Bajt,
  A.~Barty, C.~Bostedt, J.~Bozek, C.~Caleman, R.~Coffee, N.~Coppola, T.~Delmas,
  S.~Epp, B.~Erk, L.~Foucar, T.~Gorkhover, L.~Gumprecht, A.~Hartmann,
  R.~Hartmann, G.~Hauser, P.~Holl, A.~H{\"o}mke, N.~Kimmel, F.~Krasniqi, K.-U.
  K{\"u}hnel, J.~Maurer, M.~Messerschmidt, R.~Moshammer, C.~Reich, B.~Rudek,
  R.~Santra, I.~Schlichting, C.~Schmidt, S.~Schorb, J.~Schulz, H.~Soltau,
  J.~C.~H. Spence, D.~Starodub, L.~Str{\"u}der, J.~Th{\o}gersen, M.~J.~J.
  Vrakking, G.~Weidenspointner, T.~A. White, C.~Wunderer, G.~Meijer,
  J.~Ullrich, H.~Stapelfeldt, D.~Rolles, H.~N. Chapman,
  \href{https://dx.doi.org/10.1103/PhysRevLett.112.083002}{X-ray diffraction
  from isolated and strongly aligned gas-phase molecules with a free-electron
  laser}, Phys.\ Rev.\ Lett. 112 (2014) 083002.
\newblock \href {http://arxiv.org/abs/1307.4577} {\path{arXiv:1307.4577}},
  \href {http://dx.doi.org/10.1103/PhysRevLett.112.083002}
  {\path{doi:10.1103/PhysRevLett.112.083002}}.
\newline\urlprefix\url{https://dx.doi.org/10.1103/PhysRevLett.112.083002}

\bibitem{Wang:JAAP111:1}
X.-J. Wang, J.-Z. You,
  \href{http://linkinghub.elsevier.com/retrieve/pii/S016523701400374X}{Study on
  the molecular structure and thermal stability of purine nucleoside analogs},
  J.\ Anal.\ Appl.\ Pyrolysis 111 (2015) 1--14.
\newblock \href {http://dx.doi.org/10.1016/j.jaap.2014.12.024}
  {\path{doi:10.1016/j.jaap.2014.12.024}}.
\newline\urlprefix\url{http://linkinghub.elsevier.com/retrieve/pii/S016523701400374X}

\bibitem{Yablokov:RJGC:79:1704}
V.~Y. Yablokov, I.~L. Smel{\textquoteright}tsova, I.~A. Zelyaev, S.~V.
  Mitrofanova,
  \href{http://link.springer.com/10.1134/S1070363209080209}{Studies of the
  rates of thermal decomposition of glycine, alanine, and serine}, Russ. J.
  Gen. Chem. 79~(8) (2009) 1704--1706.
\newblock \href {http://dx.doi.org/10.1134/S1070363209080209}
  {\path{doi:10.1134/S1070363209080209}}.
\newline\urlprefix\url{http://link.springer.com/10.1134/S1070363209080209}

\bibitem{Zinovev:AnalChem79:8232}
A.~V. Zinovev, I.~V. Veryovkin, J.~F. Moore, M.~J. Pellin,
  \href{http://pubs.acs.org/doi/abs/10.1021/ac070584o}{Laser-driven acoustic
  desorption of organic molecules from back-irradiated solid foils}, Anal.\
  Chem. 79~(21) (2007) 8232--8241.
\newblock \href {http://dx.doi.org/10.1021/ac070584o}
  {\path{doi:10.1021/ac070584o}}.
\newline\urlprefix\url{http://pubs.acs.org/doi/abs/10.1021/ac070584o}

\bibitem{DePonte:JPD41:195505}
D.~P. DePonte, U.~Weierstall, K.~Schmidt, J.~Warner, D.~Starodub, J.~C.~H.
  Spence, R.~B. Doak,
  \href{http://iopscience.iop.org/0022-3727/41/19/195505}{{Gas dynamic virtual
  nozzle for generation of microscopic droplet streams}}, J.\ Phys.\ D 41~(19)
  (2008) 195505.
\newblock \href {http://dx.doi.org/10.1088/0022-3727/41/19/195505}
  {\path{doi:10.1088/0022-3727/41/19/195505}}.
\newline\urlprefix\url{http://iopscience.iop.org/0022-3727/41/19/195505}

\bibitem{Beyerlein:RSI86:125104}
K.~R. Beyerlein, L.~Adriano, M.~Heymann, R.~Kirian, J.~Knoska, F.~Wilde, H.~N.
  Chapman, S.~Bajt, \href{http://dx.doi.org/10.1063/1.4936843}{Ceramic
  micro-injection molded nozzles for serial femtosecond crystallography sample
  delivery}, Rev.\ Sci.\ Instrum. 86~(12) (2015) 125104--12.
\newblock \href {http://dx.doi.org/10.1063/1.4936843}
  {\path{doi:10.1063/1.4936843}}.
\newline\urlprefix\url{http://dx.doi.org/10.1063/1.4936843}

\bibitem{NIST:webbook:2017}
P.~J. Linstrom, W.~G. Mallard (Eds.), \href{http://webbook.nist.gov}{{NIST}
  Chemistry WebBook, {NIST} {S}tandard {R}eference {D}atabase {N}umber 69},
  National Institute of Standards and Technology, Gaithersburg MD, 20899, 2017.
\newblock \href {http://dx.doi.org/10.18434/T4D303}
  {\path{doi:10.18434/T4D303}}.
\newline\urlprefix\url{http://webbook.nist.gov}

\bibitem{Bondi:JAP37:4643}
A.~Bondi, \href{http://aip.scitation.org/doi/10.1063/1.1708111}{Thermal
  properties of molecular crystals. {I}. {H}eat capacity and thermal
  expansion}, J.\ Appl.\ Phys. 37~(13) (1966) 4643--4647.
\newblock \href {http://dx.doi.org/10.1063/1.1708111}
  {\path{doi:10.1063/1.1708111}}.
\newline\urlprefix\url{http://aip.scitation.org/doi/10.1063/1.1708111}

\bibitem{Hutzler:CR112:4803}
N.~R. Hutzler, H.-I. Lu, J.~M. Doyle,
  \href{http://pubs.acs.org/doi/abs/10.1021/cr200362u}{The buffer gas beam: An
  intense, cold, and slow source for atoms and molecules}, Chem.\ Rev. 112~(9)
  (2012) 4803--4827.
\newblock \href {http://dx.doi.org/10.1021/cr200362u}
  {\path{doi:10.1021/cr200362u}}.
\newline\urlprefix\url{http://pubs.acs.org/doi/abs/10.1021/cr200362u}

\bibitem{Teschmit:JCP147:144204}
N.~Teschmit, K.~D{\l}ugo{\l}\k{e}cki, D.~Gusa, I.~Rubinsky, D.~A. Horke,
  J.~K{\"u}pper, \href{http://dx.doi.org/10.1063/1.4991639}{Characterizing and
  optimizing a laser-desorption molecular beam source}, J.\ Chem.\ Phys. 147
  (2017) 144204.
\newblock \href {http://arxiv.org/abs/1706.04083} {\path{arXiv:1706.04083}},
  \href {http://dx.doi.org/10.1063/1.4991639} {\path{doi:10.1063/1.4991639}}.
\newline\urlprefix\url{http://dx.doi.org/10.1063/1.4991639}

\bibitem{Chang:IRPC34:557}
Y.-P. Chang, D.~A. Horke, S.~Trippel, J.~K{\"u}pper,
  \href{http://dx.doi.org/10.1080/0144235X.2015.1077838}{Spatially-controlled
  complex molecules and their applications}, Int.\ Rev.\ Phys.\ Chem. 34 (2015)
  557--590.
\newblock \href {http://arxiv.org/abs/1505.05632} {\path{arXiv:1505.05632}},
  \href {http://dx.doi.org/10.1080/0144235X.2015.1077838}
  {\path{doi:10.1080/0144235X.2015.1077838}}.
\newline\urlprefix\url{http://dx.doi.org/10.1080/0144235X.2015.1077838}

\bibitem{Teschmit:ACIE57:13775}
N.~Teschmit, D.~A. Horke, J.~Küpper,
  \href{https://onlinelibrary.wiley.com/doi/abs/10.1002/anie.201807646}{Spatially
  separating the conformers of the dipeptide {Ac-Phe-Cys-NH$_2$}}, Angew.\
  Chem.\ Int.\ Ed. 57~(42) (2018) 13775--13779.
\newblock \href {http://arxiv.org/abs/1805.12396} {\path{arXiv:1805.12396}},
  \href {http://dx.doi.org/10.1002/anie.201807646}
  {\path{doi:10.1002/anie.201807646}}.
\newline\urlprefix\url{https://onlinelibrary.wiley.com/doi/abs/10.1002/anie.201807646}

\end{thebibliography}
\bibliographystyle{elsarticle-num}
\end{document}